\documentclass[prb,twocolumn,showpacs]{revtex4}
\usepackage[dvips]{graphicx}
\def\b{\begin{equation}}
\def\e{\end{equation}}
\def\ba{\begin{eqnarray}}
\def\ea{\end{eqnarray}}

\begin{document}

\title{Josephson effect in mesoscopic graphene strips with finite width}
\author{Ali G. Moghaddam and Malek Zareyan}

\affiliation{Institute for Advanced Studies in Basic Sciences,
45195-1159, Zanjan, Iran}

\begin{abstract}
We study Josephson effect in a ballistic graphene strip of length
$L$ smaller than the superconducting coherence length and
arbitrary width $W$.  We find that the dependence of the critical
supercurrent $I_{c}$ on $W$ is drastically different for different
types of the edges. For \textit{smooth} and \textit{armchair}
edges at low concentration of the carriers $I_{c}$ decreases
monotonically with decreasing $W/L$ and tends to a constant
minimum for a narrow strip $W/L\lesssim1$. The minimum
supercurrent is zero for smooth edges but has a finite value
$e\Delta_{0}/\hbar$ for the armchair edges. At higher
concentration of the carriers, in addition to this overall
monotonic variation, the critical current undergoes a series of
peaks with varying $W$. On the other hand in a strip with
\textit{zigzag} edges the supercurrent is half-integer quantized
to $(n+1/2)4e\Delta_{0}/\hbar$, showing a step-wise variation with
$W$.
\end{abstract}





\pacs{74.45.+c, 74.50.+r, 73.63.-b, 74.78.Na} \maketitle

Over the past years fabrication of the nanoelectronic structures
has provided the possibility of detecting the effects of the
electronic transport through a few or even single quantum states.
The generic effect is the quantization of the conductance of a
quantum point contact (QPC) to $2e^2/\hbar$ times the number of
the transparent quantum channels \cite{vanwees88,wharam88}. In a
superconducting quantum point contact (SQPC) the analogous
behavior occurs for the Josephson current \cite{beenakker91-1}.
Furusaki \textit{et al.} \cite{furusaki91} have predicted that the
critical (maximum) supercurrent through a two-dimensional electron
gas (2DEG) junction shows a step-like variation as a function of
the width of the junction depending on the Fermi wave length of
the electrons. The quantization of the supercurrent in units of
$e\Delta_{0}/\hbar$ ($\Delta_{0}$ is the superconducting gap in
the electrodes) was confirmed experimentally in semiconducting
structures \cite{takayanagi95}.

Recently experimental realization of graphene, the two dimensional
carbon atoms arranged in a honeycomb lattice, has introduced a new
type of mesoscopic material with unique properties
\cite{novoselov04,novoselov05,zhang05}. Most of the peculiarities
come from the electronic structure of graphene which is
fundamentally different from that of a metal or a semiconductor.
Graphene has a gapless semi-metallic band structure with a linear
dispersion relation of the low-lying excitations. This makes the
electrons in graphene to behave identical to two dimensional
massless Dirac Fermions\cite{wallace,slonczewski,semenoff}.
Already several quantum transport phenomena including the integer
quantum Hall effect \cite{novoselov05,zhang05,gusynin05}, the
conductance quantization\cite{peres06} and quantum shot noise
\cite{beenakker06-1} have been revisited in graphene and found to
have anomalous features due to the relativistic-like dynamic of
the electrons.

Very interestingly graphene as a weak link between two
superconductors can serve as a relativistic Josephson contact
\cite{beenakker06-2}. Together with the fact that the graphene
samples are realized with different type of the edges suggest a
new class of Josephson devices with novel properties. In
particular the quantization of the supercurrent through a graphene
strip with finite width is expected to depend on the type of the
edges. The aim of this paper is to study graphene Josephson
junctions with different edges.

Josephson effect in a wide graphene sheet contacting two
superconductors was studied by Titov and
Beenakker\cite{beenakker06-2} just recently. Using a
Dirac-Bogoliubov-de Gennes (DBdG) formalism \cite{beenakker06-3},
they found that in a ballistic graphene a Josephson current can
flow even in the limit of zero concentration of the carriers
\textit{i.e.} at the Dirac point. The dependence of the critical
current on the junction length as well as the current-phase
relation were found to be the same as in a Josephson junction with
an ordinary \textit{disordered} normal metal
\cite{beenakker91-2,kulik1}.

In this paper using the formalism of
Ref.\onlinecite{beenakker06-3} we study the Josephson effect in a
graphene strip with length $L$ smaller than the superconducting
coherence length $\xi=\hbar v /\Delta_{0}$ and an arbitrary width
$W$ for smooth, armchair and zigzag edges. We find that in
contrast to an ordinary SQPC \cite{beenakker91-1,furusaki91} the
supercurrent in smooth and armchair strips with a low
concentration of the carriers is not quantized but rather shows a
monotonic decrease with lowering $W$. For a narrow strip $
W\lesssim L$ the supercurrent reduces to a constant minimum which
is $e\Delta_{0}/\hbar$ for armchair edges and vanishingly small
for smooth edges. Far from the Dirac point at a high concentration
of the carriers the behavior of the supercurrent versus $W$
consists of a similar monotonic variation and a series of peaks
with distances inversely proportional to the chemical potential
$\mu$. We explain absence of the supercurrent quantization as a
result of the significant contribution of the evanescent modes in
the supercurrent which is a unique property of graphene
monolayers. For a zigzag strip the situation is drastically
different because of the valley filtering characteristic
\cite{rycerz06} of the wave functions. We find that this type of
the edges supports a half-integer quantization of the supercurrent
to $(n+1/2)4e\Delta_{0}/\hbar$.

\begin{figure}
\centerline{\includegraphics[width=8cm,angle=0]{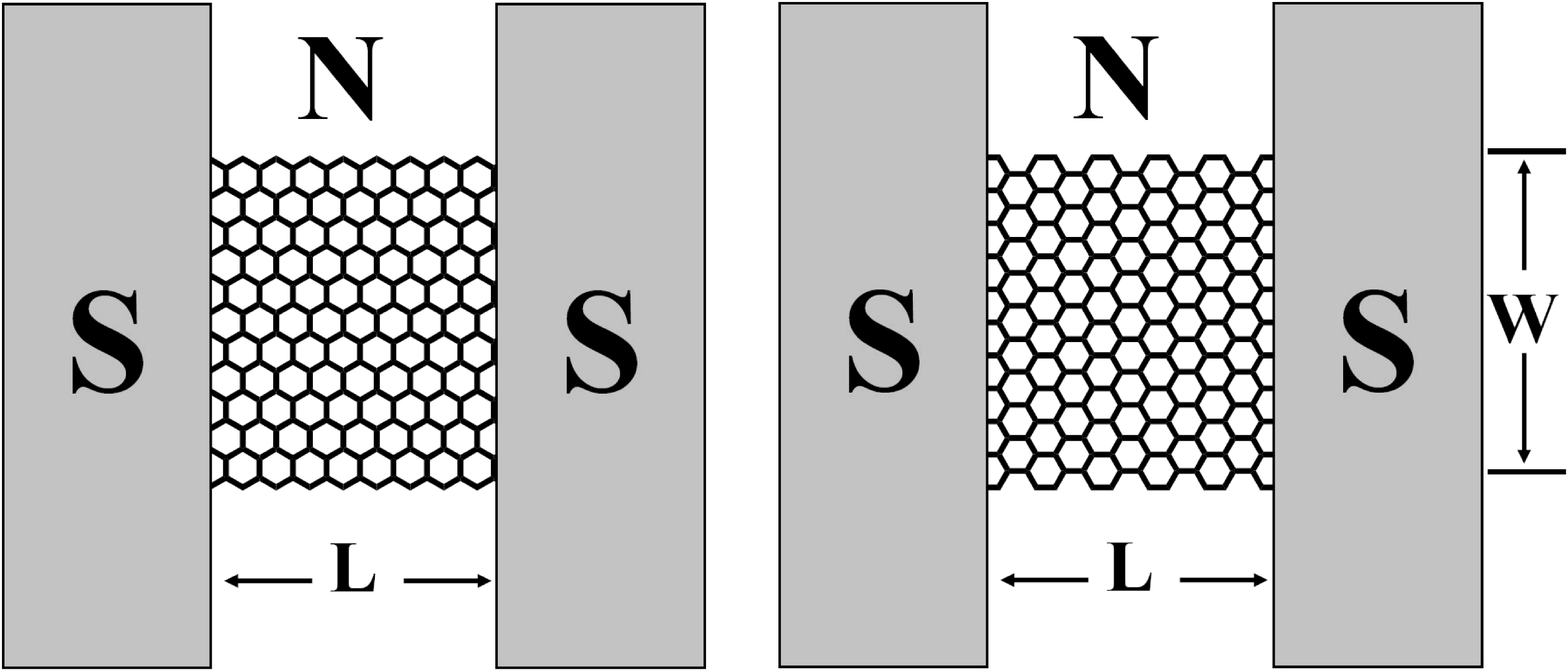}}
\caption{Schematic of the studied junction structures. Right
(left) junction has armchair (zigzag) edges.} \label{gzfig1}
\end{figure}

The geometry of the studied Joseohson junction is shown
schematically in Fig. \ref{gzfig1}. A ballistic graphene strip (N)
connects two wide superconducting regions (S) which are produced
by depositing of superconducting electrods on top of the graphene
sheets. We assume the interfaces between the superconductors and
the graphene strip to be free of defects and impurities. The
superconducting parts are assumed to be heavily doped such that
the Fermi wave length $\lambda_{F\,S}$ inside them is very smaller
than the superconducting coherence length $\xi$ and also Fermi
wave length of the normal graphene strip $\lambda_{F\,N}$. By the
first condition mean field theory of superconductivity will be
justified and by second we can neglect the spatial variation of
$\Delta$ in the superconductors close to the normal-superconductor
(NS) interfaces. Thus the superconducting order parameter has the
constant values $\Delta=\Delta_{0}\exp(\pm i\phi/2)$ in the left
and right superconductor, respectively, and vanishes identically
in N.

\par
The superconducting correlations between electron-hole excitations
are described by DBdG equation for the electron and hole wave
functions $u$ and $v$ \cite{beenakker06-3}:

\begin{equation}
\left(\matrix{\hat{H}-\varepsilon-\mu & \hat{\Delta} \cr
\hat{\Delta}^{*} & \mu-\hat{\Theta} \hat{H}\hat{\Theta}
^{-1}-\varepsilon}\right)\left(\matrix{u\cr v}\right) = 0.
\label{dbdg}
\end{equation}
Here $\varepsilon$ is the excitation energy and the single
particle Dirac Hamiltonian $\hat{H}=-i\,\hbar
v(\partial_{x}\,\sigma_{1}\,\tau_{0}+\partial_{y}\,\sigma_{2}\,\tau_{3})$,
the time-reversal operator $\hat{\Theta}=\sigma_{3}\,\tau_{1}$,
and the superconductivity pair potential
$\hat{\Delta}=\Delta\,\sigma_{0}\,\tau_{1}$, operate on the wave
functions $u$ and $v$. The wave functions are 4-component spinors
$(\psi_{1},\psi_{2},\psi'_{1},\psi'_{2})$ in which the indices
show the two sublattices of the honeycomb lattice and primed
(unprimed) $\psi$ indicates $K^{\prime}$ ($K$) Dirac point in
inverse honeycomb lattice. $\sigma_{i}$ and $\tau_{i}$
($i=1,2,3$), are Pauli matrices in the sublattice and valley
spaces respectively with the corresponding unit matrices
$\sigma_{0}$ and $\tau_{0}$. Eq. \ref{dbdg} describes the
superconducting correlations between massless Dirac electron and
hole excitations with opposite spins and different valleys.
\par
The phase difference $\phi$ between the order parameters drives a
supercurrent through the graphene strip which constitutes a weak
link between two superconductors. This Josephson supercurrent is
carried by the so called Andreev states which are formed in N
region due to the successive conversion of the electron-hole
excitations to each other (Andreev reflection) at the NS
boundaries. For a short junction of $L\ll\xi$ and at zero
temperature the Andreev bound (discrete) states with energies
$|\varepsilon|<\Delta_{0}$ have the main contribution to the
Josephson supercurrent \cite{beenakker91-1}. In this case we can
neglect the contribution from the continuous states above the
superconducting gap. We obtain the energies $\varepsilon(\phi)$ of
Andreev bound states by solving the DBdG equations with
appropriate boundary conditions which are described below. The
Josephson current is then obtained from the formula

\begin{equation}
I=\frac{4e}{\hbar}\sum_{n} \frac{d}{d\phi}\varepsilon_{n}(\phi),\\
\label{i-phi}
\end{equation}
where the factor $4$ accounts for the spin and valley
degeneracies.
\par
Inside N the solutions of DBdG equations are the electron and
hole-like wave functions which are classified by the 2-dimensional
wave vector ${\mathbf{k}}\equiv(k,q)$ with the energy-momentum
relation $\varepsilon =\hbar v|\mathbf{k}|$. For a finite
width $W$ the transverse momentum is quantized ($q_{n}$) by
imposing the boundary conditions at the edges. The transverse
boundary conditions are different for different edge types which a
graphene strip can have.
\par
The solutions inside the S are rather mixed electron-hole
excitations, called Dirac-Bogoliubov quasiparticles.  In general
to match the solutions inside different regions one needs the
scattering matrices at the interfaces. But assuming ideal NS
boundaries electrone-hole scattering can be described by a
longitudinal boundary condition between electron and hole wave
functions which read \cite{beenakker06-2},
\begin{equation}
u=e^{-i\varphi+i\beta\mathbf{n}.\mathbf{\sigma}}\,\tau_{1} v,\\
\,\,\,\,\beta=\arccos(\varepsilon/\Delta_{0})\\
\label{e-h}
\end{equation}
where $\varphi$ is the phase of $\Delta$ in S and $\mathbf{n}$ is
the unit vector perpendicular to the interface pointing from N to
S.
\par
For each type of the edges by imposing the longitudinal boundary
conditions at the interfaces we obtain the Andreev energies
$\varepsilon_n(\phi)$ of the transverse mode $n$. We will consider
the regime $\Delta_{0}\ll\mu$ where the retro-Andreev reflection
dominates\cite{beenakker06-3}. In this limit the longitudinal
momentums of electron and hole $k_{n}(\pm \varepsilon)$ can be
approximated with $k_{n}(0)$ .
\par
Let us start with analyzing a graphene strip with smooth edges
which corresponds to an infinite mass confinement. For this case
the transverse boundary conditions do not mix the valleys and have
the form $\psi_{1}(0) = -\psi_{2}(0)$ and $\psi_{1}(W) =
\psi_{2}(W)$ \cite{berry}. For the electron-like Andreev bound
state the two-component wave function is a composition of the
right and left going waves $u(k)$ and $u(-k)$ with
$u(k)=e^{ikx}[\sin(qy-\alpha/2),\sin(qy+\alpha/2)]$, and a
quantized transverse momentum $q_{n}=(n+\frac{1}{2})\pi/W$,
($n=0,1,2,...$). The angle $\alpha=\arctan(q/k)$ indicates the
propagation direction. Correspondingly the hole-like wave
functions are written as a composition of $v(k)$ and $v(-k)$ which
are the time-reversed of the electron wave functions. Imposing the
electron-hole conversion condition (Eq. \ref{e-h}) at the NS
boundaries ($x=0, L$) we obtain the energy of the Andreev state as

\begin{equation}
\varepsilon_{n}(\phi)=\Delta_{0}\sqrt{1-t_{n}\sin^{2}(\phi/2)},\\
\label{e-phi}
\end{equation}
where $t_{n}=[1+(q_{n}/k_{n})^{2}\sin^{2}(k_{n}L)]^{-1}$, is the
transmission coefficient of the mode $n$ through the junction.
\par
For the armchair edges the two valleys are mixed by the boundary
conditions $\psi_{1}=\psi'_{1}$ and $\psi_{2}=\psi'_{2}$ at the
two edges \cite{brey06}. The electron wave function is a mixture
of the two valleys with the form

\begin{equation}
u=e^{ikx}[e^{iqy}(e^{-i\alpha},e^{i\alpha}),
(-1)^{n}e^{-iqy}(e^{-i\alpha},e^{i\alpha})],\\
\end{equation}
where the transverse momentum quantized to $q_{n}=n\pi/W$ with
integer $n$. Note that in contrast to the smooth edges, the wave
functions $u(q)$ and $u(-q)$ represent distinct solutions in which
the components of the two valleys are interchanged. As a result
the lowest mode $n=0$ has not the two fold valley degeneracy of
the higher $n=1,2,...$ modes. Like the case of smooth edges the
electron wave function of a given mode $n$ is the composition of
$u(k)$ and $u(-k)$. We find that the Andreev energies have the
same form as in the smooth strip(Eq. \ref{e-phi}).
\par
We now consider the case of zigzag edges. In contrast to the
smooth and armchair edges, these have a property of coupling the
longitudinal and transverse momentums $k$ and $q$. The boundary
conditions are $\psi_{1}=\psi'_{1}=0$ at $y=0$ and
$\psi_{2}=\psi'_{2}=0$ at $y=W$ \cite{brey06}, which lead to the
transcendental relation $\sin(qW)/qW=\pm \hbar v /\mu W $ for the
transverse momentum $q$. This relation has a finite number of
solutions depending on the value of $\mu W/\hbar v$. For $\mu
W/\hbar v<1$ there is an imaginary solution $q$ labeling an
evanescent mode in the $y$-direction.  In the interval $1<\mu
W/\hbar v<3\pi/2$, the transcendental relation has a single
oscillatory solution. For larger $\mu W/\hbar v$ the number of the
solutions are increased by two whenever the width is increased by
a half of Fermi wave length $\lambda_{F\,N}=hv/\mu$. We classify
the wave functions to two groups according to $\pm$ sings in the
transcendental relation. The electronic wave functions of the
first group ($+$ sign) have the forms

\begin{equation}
u^{+}=e^{ikx}[\sin(qy),\sin(qy+\alpha),0,0],\\
\end{equation}
\begin{equation}
u^{-}=e^{-ikx}[0,0,\sin(qy),\sin(qy-\alpha)],\\
\end{equation}
which, respectively, describe a right-going wave on the valley $K$
and a left-going wave in the other valley $K^{\prime}$. The wave
functions of the second group ($-$ sign) is obtained from the
above functions by the replacement $k\rightarrow-k$. We note that
for each mode the zigzag strip operates as a valley filter for the
waves with mixed valley components \cite{rycerz06}. Since the
valley index and $q$ are conserved upon a normal reflection, the
filtering property prevents the electrons from such scatterings.
Consequently the wave functions of the electron bound states in N
will be a composition of $u^{+ (-)}$ and $v^{-(+)}$. From the
longitudinal condition (Eq. \ref{e-h}) we obtain that the energy
of the Andreev state does not depend on the transverse mode $q$
and is given by $\varepsilon=\Delta_{0}\cos(\phi/2)$, which is the
same as for an ordinary SQPC \cite{beenakker91-1}.

\begin{figure}
\centerline{\includegraphics[width=8cm,angle=0]{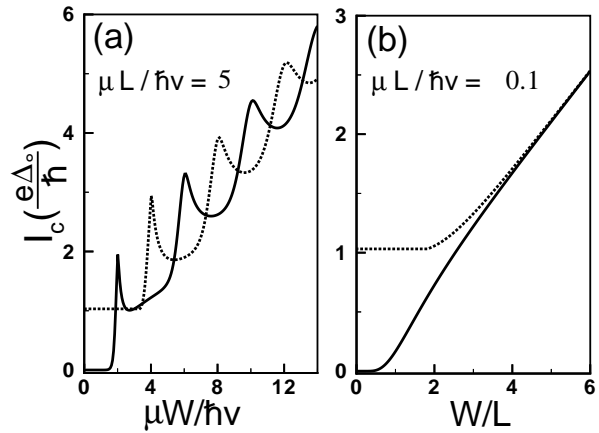}}
\caption{The critical current \textit {vs.} relative width of
junction for two different values of $\mu L/\hbar v$. Solid
(dashed) line is corresponding to junction with smooth (armchair)
edges.} \label{gzfig2}
\end{figure}
\par
From Eq. \ref{i-phi} and using the above obtained results for the
allowed transverse momentums $q$ and the Andreev energies
$\varepsilon$ we obtain the Josephson current for the three edge
types. The results of smooth and armchair edges are similar. In
both cases there are infinite numbers of transverse modes. For a
given $\mu L/\hbar v$ there are oscillatory modes with real $k$ as
well as evanescent modes with imaginary $k$. Unlike the ordinary
QPCs the evanescent modes in the graphene strip may have a
significant transmission depending on the value of $\mu L/\hbar
v$. In particular at the Dirac point $\mu L/\hbar v\ll 1$ there is
only few propagating modes and the evanescent modes have the main
contribution in the supercurrent. This is the unique property of
the graphene Josephson junctions in which very small Fermi
energies are accessible.
\par
Fig. \ref{gzfig2} shows the critical current dependence on width
for the smooth and armchair edges and for two values of $\mu
L/\hbar v$. For a typical small value $\mu L/\hbar v=0.1$ (Fig.
\ref{gzfig2}b) $I_{c}$ has a linear dependence on $W/L$ at large
$W/L$ which is the result of a diffusive-like transport in the
ballistic graphene \cite{beenakker06-2}. By decreasing $W/L$ the
transmission of the modes through N decreases and the critical
current shows a monotonic decrease without any quantization. For a
narrow strip $W\lesssim L$ the evanescent modes transparencies is
vanishingly small and $I_{c}$ reduces to a constant minimum value.
For the smooth edges no propagating mode can exist when
$W<(\pi/2)\hbar v/\mu$. So the minimum supercurrent goes to zero.
However in the case of the armchair edges a lowest nondegenerate
mode with $q=0$ always can propagate irrespective of the width of
the junction. This zero mode result in a bound state carrying a
nonzero residual supercurrent $I_{c}=e\Delta_{0}/\hbar$ for
$W\lesssim L$ (see Fig. \ref{gzfig2}b).
\par
For a higher chemical potential, $\mu L/\hbar v=5$ (Fig.
\ref{gzfig2}a) an oscillatory variation is added to the monotonic
behavior. In this regime both the propagating and evanescent modes
contribute to the supercurrent. In addition to an overall increase
with $\mu W/\hbar v$ the critical current undergoes peaks which
become smoother by increasing $\mu W/\hbar v$. Each peak (jump)
signals addition of a new propagating mode in the transport. Note
that due to a nonzero transverse momentum of the lowest mode
($n=0$) in smooth-edge strip the oscillations are shifted by half
a period with respect to those of the armchair-edge one. Overall
absence of the supercurrent quantization in these junctions is a
consequence of the existence of the evanescent modes with
appreciable transmission through N.
\par
\begin{figure}
\centerline{\includegraphics[width=8cm,angle=0]{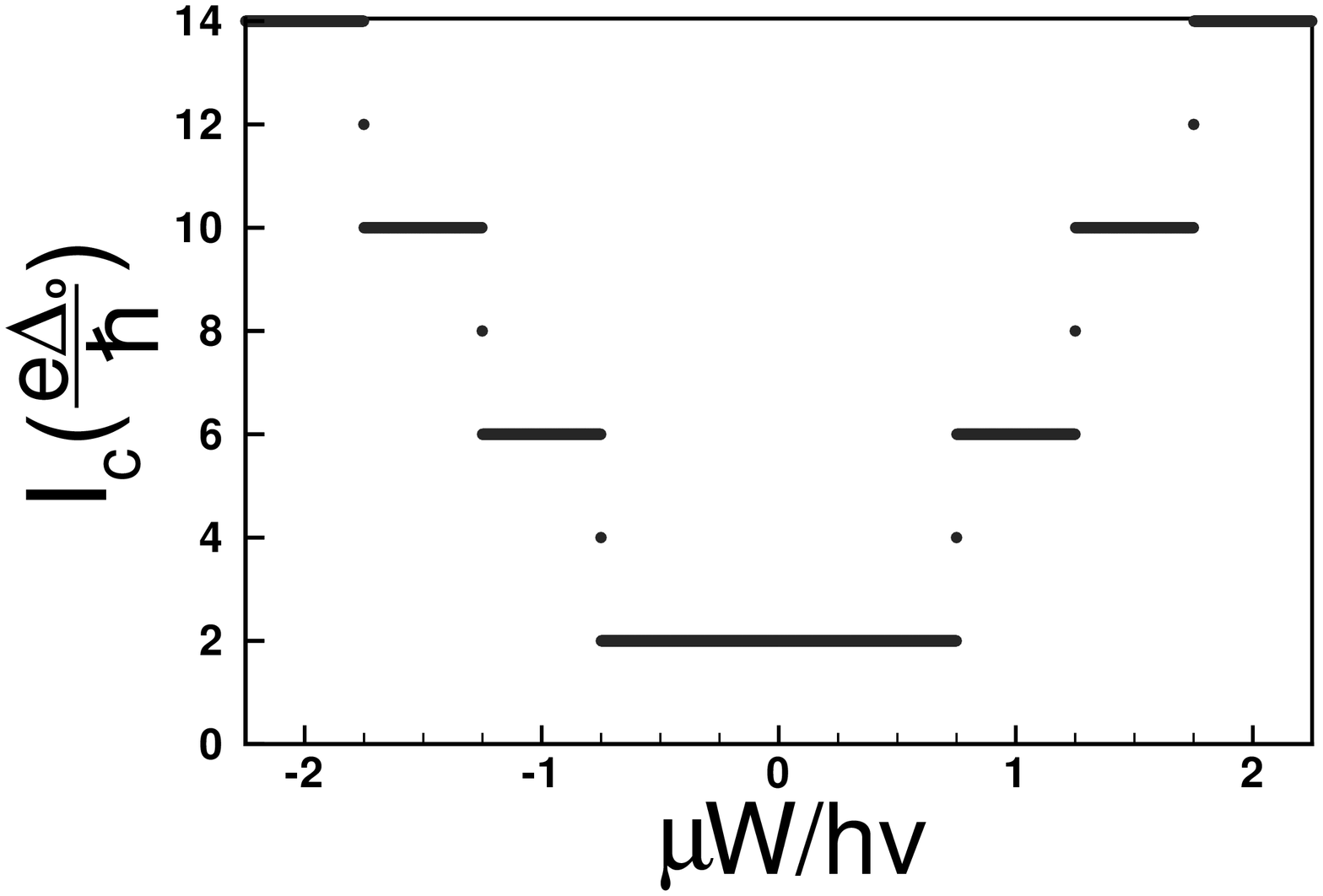}}
\caption{The critical current \textit {vs.} relative width of
junction with zigzag edges.} \label{gzfig3}
\end{figure}

Finally let us analyze the Josephson supercurrent for the zigzag
strip. As we argued above in this case the valley filtering nature
of the wave functions in N prevents the normal reflections at
NS-interfaces. There is no evanescent mode with imaginary $k$ and
the situation is similar to an ordinary SQPC.  The critical
supercurrent shows a step-wise variation with $\mu W/hv$ but with
the following important anomalies (see Fig. \ref{gzfig3}). In
contrast to an ordinary SQPC \cite{beenakker91-1,furusaki91}, the
width of the first step ($3/4$) is bigger than that of the higher
steps ($1/2$). The extra width is the contribution of the single
$y$-direction evanescent mode (imaginary $q$) for $W<\hbar v/\mu
$. Also the height of the first step is $1/2$ of the height of the
higher steps which itself is four times bigger than
$e\Delta_{0}/\hbar$ the supercurrent quanta in an ordinary SQPC.
Therefore the supercurrent through a zigzag graphene strip is
half-integer quantized to $(n+1/2)4e\Delta_{0}/\hbar$. The effect
resembles the conductance quantization in graphene strips with
zigzag edges \cite{peres06} and also the half-integer Hall effect
\cite{novoselov05,zhang05,gusynin05} in monolayer structures.

In conclusion we have investigated the Josephson effect in a short
monolayer graphene strip connecting two superconducting
electrodes. Within a Dirac-Bogoliubov-de Gennes formalism, we have
found that the variation of the critical supercurrent $I_c$ versus
the width $W$ is drastically different for the graphene strips
with different edges. In the smooth and armchair strips with low
concentration of the carriers $I_c$ decreases monotonically with
decreasing $W/L$. For a narrow strip $W\lesssim L$, $I_c$ takes a
constant minimum value which is given by $e\Delta_{0}/\hbar$ for
the armchair edges but vanishingly small for the smooth edges.
Increasing the concentration of the carriers, this overall
monotonic dependence of $I_c$ acquires a series of peaks with
distances inversely proportional to the chemical potential $\mu$.
For the zigzag edges we have found a step-wise variation of $I_c$
versus $\mu W/hv$, implying a half-integer quantization of the
supercurrent to $(n+1/2)4e\Delta_{0}/\hbar$ with integer $n$.


\begin{thebibliography}{99}

\bibitem{vanwees88}
B. J. van Wees, H. van Houten, C. W. J. Beenakker, J. G.
Williamson, L. P. Kouwenhoven, D. van der Marel, and C. T. Foxon
Phys. Rev. Lett. {\bf 60}, 848 (1988).

\bibitem{wharam88}
A. Wharam, T. J. Thornton, R. Newbury, M. Pepper, H. Ahmed, J. E.
Frost, D. G. Hasko, D. C. Peacock, D. A. Richie, and G. A. C.
Jones, J. Phys. C {\bf 21}, L209 (1988).

\bibitem{beenakker91-1}
C. W. J. Beenakker and H. van Houten Phys. Rev. Lett. {\bf 66},
3056 (1991).

\bibitem{furusaki91}
A. Furusaki, H. Takayanagi, and M. Tsukada, Phys. Rev. Lett.
{\bf 67}, 132 (1991); Phys. Rev. B {\bf 45}, 10563 (1992).

\bibitem{takayanagi95}
H. Takayanagi, T. Akazaki, and J. Nitta, Phys. Rev. Lett.
{\bf 75}, 3533 (1995).

\bibitem{novoselov04}
K.S. Novoselov, A.K. Geim, S.V. Morozov, D. Jiang, Y. Zhang, S.V.
Dubonos, I.V. Grigorieva, and A.A. Firsov, Science {\bf 306}, 666
(2004).

\bibitem{novoselov05}
K. S. Novoselov, A. K. Geim, S.V. Morozov, D. Jiang, M.I.
Katsnelson, I.V. Grigorieva, S.V. Dubonos, and A.A. Firsov, Nature
{\bf 438}, 197 (2005).

\bibitem{zhang05}
Y. Zhang, Y. W. Tan, H. L. Stormer, and P. Kim, Nature {\bf 438},
201 (2005).

\bibitem{wallace}
P. R. Wallace Phys. Rev. {\bf 71}, 622 (1947).

\bibitem{slonczewski}
J. C. Slonczewski and P. R. Weiss, Phys. Rev. {\bf 109}, 272
(1958).

\bibitem{semenoff}
G. W. Semenoff, Phys. Rev. Lett. {\bf 53}, 2449 (1984).

\bibitem{gusynin05}
V. P. Gusynin and S. G. Sharapov,
Phys. Rev. Lett. {\bf 95}, 146801 (2005).

\bibitem{peres06} N. M. R. Peres, A. H. Castro Neto, and F. Guinea,
Phys. Rev. B {\bf 73}, 195411 (2006).


\bibitem{beenakker06-1}
J. Tworzydlo, B. Trauzettel, M. Titov, A. Rycerz, and C.W.J.
Beenakker, Phys. Rev. Lett. {\bf 96}, 246802 (2006).

\bibitem{beenakker06-2}
M. Titov and C. W. J. Beenakker, Phys. Rev. B {\bf 74} , 041401(R)
(2006).

\bibitem{beenakker06-3}
C. W. J. Beenakker, Phys. Rev. Lett. {\bf 97}, 067007 (2006).

\bibitem{beenakker91-2}
C. W. J. Beenakker, Phys. Rev. Lett. {\bf 67}, 3836 (1991); {\bf
68}, 1442(E) (1992).

\bibitem{kulik1}
I. O. Kulik and A. N. Omelyanchuk, Pisma Zh. Eksp. Teor. Fiz.
{\bf 21}, 216 (1975); JETP Lett. {\bf 21}, 96 (1975).

\bibitem{rycerz06}
A. Rycerz, J. Tworzydlo, C. W. J. Beenakker, cond-mat/0608533

\bibitem{berry}
M. V. Berry and R. J. Mondragon, Proc. R. Soc. Lond. A {\bf 412},
53 (1987).

\bibitem{brey06}
L. Brey and H. A. Fertig, Phys. Rev. B {\bf 73}, 235411 (2006).
\end{thebibliography}
\end{document}